\documentclass[sigconf]{acmart}
\let\Bbbk\relax 
\Bbbk\usepackage{amsmath,amssymb,amsfonts}
\usepackage{algorithmic}
\usepackage{algorithm}
\usepackage{array}
\usepackage{booktabs}
\usepackage{color}
\usepackage{graphicx}
\usepackage{multirow}
\usepackage{bm}
\usepackage{url}
\usepackage{enumitem}
\usepackage{titlesec}

\titlespacing*{\subsubsection}{0pt}{0pt}{5pt}

\AtBeginDocument{%
  }

\setcopyright{acmlicensed}
\copyrightyear{2025}
\acmYear{2025}
\acmDOI{XXXXXXX.XXXXXXX}
\acmConference[Conference acronym 'XX]{Make sure to enter the correct
  conference title from your rights confirmation email}{June 03--05,
  2018}{Woodstock, NY}
\acmISBN{978-1-4503-XXXX-X/2018/06}

\copyrightyear{2025} 
\acmYear{2025} 
\setcopyright{cc}
\setcctype{by-nc-sa}
\acmConference[aiDM '25]{International Workshop on Exploiting Artificial Intelligence Techniques for Data Management}{June 22--27, 2025}{Berlin, Germany}
\acmBooktitle{International Workshop on Exploiting Artificial Intelligence Techniques for Data Management (aiDM '25), June 22--27, 2025, Berlin, Germany}\acmDOI{10.1145/3735403.3735996}
\acmISBN{979-8-4007-1920-2/2025/06}

\begin{document}

\title{Filter-Centric Vector Indexing: Geometric Transformation for Efficient Filtered Vector Search}


\author{Alireza Heidari}
\orcid{0009-0002-0414-7360}
\affiliation{%
  \institution{Huawei}
  \city{Vancouver}
  \country{Canada}}
\email{alireza.heidarikhazaei@huawei.com}

\author{Wei Zhang}
\orcid{0009-0006-2866-8688}
\affiliation{%
  \institution{Huawei}
  \city{Vancouver}
  \country{Canada}}
  \email{wei.zhang6@huawei.com}







\begin{abstract}
The explosive growth of vector search applications demands efficient handling of combined vector similarity and attribute filtering—a challenge where current approaches force an unsatisfying choice between performance and accuracy. We introduce Filter-Centric Vector Indexing (FCVI), a novel framework that transforms this fundamental trade-off by directly encoding filter conditions into the vector space through a mathematically principled transformation $\psi(v, f, \alpha)$. Unlike specialized solutions, FCVI works with any existing vector index (HNSW, FAISS, ANNOY) while providing theoretical guarantees on accuracy. Our comprehensive evaluation demonstrates that FCVI achieves 2.6-3.0× higher throughput than state-of-the-art methods while maintaining comparable recall. More remarkably, FCVI exhibits exceptional stability under distribution shifts—maintaining consistent performance when filter patterns or vector distributions change, unlike traditional approaches that degrade significantly. This combination of performance, compatibility, and resilience positions FCVI as an immediately applicable solution for production vector search systems requiring flexible filtering capabilities.
\end{abstract}





\maketitle

\section{INTRODUCTION}
Vector databases have become critical infrastructure for modern AI applications, managing high-dimensional embeddings that represent semantic information in text, images, audio, and other modalities~\cite{chan1998bitmap,heidari2019holodetect,heidari2024record,kiela2018efficient}. Real-world applications rarely rely on pure similarity search alone - they almost always combine vector similarity with traditional filtering predicates to narrow the search space and improve relevance.

\noindent For example, an e-commerce product search might filter by price range before finding visually similar items; a recommendation system might filter by recency before matching semantic similarity. These hybrid queries pose significant technical challenges, as they require efficiently combining fundamentally different operations: vector similarity search and scalar filtering.

\subsection{Approach and Technical Challenges}
In this paper, we introduce a fundamentally different approach to filtered vector search: a geometric transformation that incorporates filter values directly into the vector space itself while preserving the original dimensionality. Our key insight is that when filter data consist of numbers or vectors with dimensions fewer than the embedding vector space, we can partition vectors to match filter dimensions and recoordinate them based on filter values.

Specifically, we identify vectors with the same filter values and consider them as a cluster with a center determined by the filter vector. We then transform the original vector space by adjusting the vectors according to their filter centers. In this new geometric space, vectors with similar filter values are brought closer together, while vectors with different filter values are pushed further apart, creating a unified space where both filter similarity and vector similarity are preserved.

Unlike existing approaches that create separate structures for vectors and filters or require complex segmentation strategies, our transformation approach directly incorporates filter values into the vector space without changing dimensionality, allowing any standard vector index to efficiently handle filtered queries.

This approach faces several technical challenges:
\begin{itemize}[leftmargin=1.5em]
\item Balancing the influence of filter values versus original vector similarity
\item Handling high-cardinality or continuous filter attributes
\item Efficiently updating the index when data changes
\item Supporting multiple filter attributes with different characteristics
\item Preserving search quality while improving performance
\end{itemize}

\subsection{Contributions and Organization}
This paper makes the following contributions:

\begin{itemize}[leftmargin=1.5em]
\item We introduce Filter-Centric Vector Indexing (FCVI), a novel approach that incorporates filter attributes directly into vector space, enabling efficient and accurate filtered vector search.
\item We develop a formal theoretical model for filter-aware vector transformation with provable guarantees on distance preservation and query accuracy.
\item We present a unified framework that outperforms traditional pre-filtering, post-filtering, and hybrid approaches, demonstrating superior stability under distribution shifts.
\item Extensive experiments show FCVI achieves 1.4-3.0$\times$ lower latency than state-of-the-art methods while maintaining high recall (94.8-95.3\%) and requiring minimal additional resources.
\end{itemize}
\noindent The remainder of this paper is organized as follows. Section \ref{sec:bg} provides background on vector search and filtering approaches. Section \ref{sec:overview} presents our overview of the FCVI framework. Section \ref{sec:filter-query} analyzes the representations of filtered queries, followed by theoretical analysis in Section \ref{sec:theory}. Section \ref{sec:exp} evaluates performance and stability under various conditions. Section \ref{sec:related} discusses related work, and Section \ref{sec:conclusion} concludes.
\section{BACKGROUND}
\label{sec:bg}
\subsection{Vector Indexing}
Vector databases store high-dimensional vectors (typically 100-1000 dimensions) that represent semantic information extracted through machine learning models~\cite{heidari2024uplif,heidari2025doblix}. Given a query vector $q$, these systems aim to efficiently find the $k$ most similar vectors according to some distance function, typically Euclidean distance or cosine similarity.

However, exact nearest neighbor search becomes computationally infeasible in high dimensions due to the "curse of dimensionality."~\cite{koppen2000curse,adc} To address this, vector databases employ approximate nearest neighbor (ANN) techniques that trade perfect recall for significant speed improvements. Popular ANN algorithms include:

\begin{itemize}[leftmargin=1.5em]
\item \textbf{Hierarchical Navigable Small World (HNSW)} \cite{malkov2018efficient}: Creates a multi-layer graph structure with logarithmic search complexity.
\item \textbf{Inverted File (IVF)} \cite{jegou2010product}: Partitions the vector space into clusters and searches only the most promising clusters.
\item \textbf{Product Quantization (PQ)} \cite{jegou2011searching}: Compresses vectors by splitting them into subvectors that are independently quantized.
\end{itemize}

These techniques rely on preprocessing the vector space to create index structures that accelerate search. However, they are optimized for pure vector similarity search and do not naturally accommodate filtering conditions.

\subsection{Filter Query}
In traditional databases, filtering involves selecting records that satisfy specific conditions on attribute values (e.g., price $< 100$, category = "animals"). These operations are typically handled efficiently through B-trees, hash indexes, or bitmap indexes.

When filtering is combined with vector search, existing approaches fall into several categories:

\begin{itemize}[leftmargin=1.5em]
\item \textbf{Post-filtering}: Perform vector search first and then apply filters to the results. This approach becomes inefficient when filters are highly selective, as many retrieved vectors may be discarded.

\item \textbf{Pre-filtering}: Apply filters first, then perform vector search on the filtered subset. This approach suffers when filters return large sets of results, as it loses the benefit of efficient ANN indexes \cite{wang2021milvus}.

\item \textbf{Hybrid filtering}: Recent work like UNIFY \cite{unify} attempts to combine pre- and post-filtering by segmenting data by attribute values and selecting the optimal strategy based on query range size, but requires maintaining complex graph structures.
\end{itemize}

\noindent Our work takes a fundamentally different approach by transforming the vector space itself to incorporate filter information, allowing a single vector index to efficiently handle filtered queries.
\section{FRAMEWORK OVERVIEW}
\label{sec:overview}
\subsection{Problem Statement}
We formally define the filtered vector search problem as follows:

\noindent Let $\mathcal{D} = \{(v_1, f_1), (v_2, f_2), ..., (v_n, f_n)\}$ be a dataset where each item consists of a vector $v_i \in \mathbb{R}^d$ and a filter vector $f_i \in \mathbb{R}^m$ with $m < d$. Each filter vector encodes one or more attributes of the item (e.g., price, category, timestamp) into a numerical representation. For categorical attributes, we use one-hot encoding or learned embeddings, while for numerical attributes, we normalize values to the range appropriate for our transformation. Multiple filter attributes are concatenated to form the final filter vector. The filter vector may represent multiple filter attributes combined into a single vector.

To ensure mathematical stability and comparability, we assume that each dimension of both vectors and filters is normalized to follow a standard normal distribution based on the values in that dimension. Specifically, for each dimension $j$ of vector $v_i$ and dimension $k$ of filter $f_i$:
\begin{align}
    & v_{i,j} \sim \mathcal{N}(0, 1) \quad \forall j \in \{1,2,...,d\} \\
    & f_{i,k} \sim \mathcal{N}(0, 1) \quad \forall k \in \{1,2,...,m\}
\end{align}

\noindent where normalization is performed independently for each dimension across the dataset.

Given a query $(q, F_q)$ consisting of a query vector $q \in \mathbb{R}^d$ and a filter query vector $F_q \in \mathbb{R}^m$, our aim is to find the $k$ vectors that optimize a combined objective of: \textbf{(I)} Maximizing filter similarity: $\text{sim}(f_i, F_q) \rightarrow 1$. \textbf{(II)} Minimizing the distance to the query vector: $\text{dist}(v_i, q)$.

Unlike traditional approaches with strict binary predicates, our formulation generalizes filter matching and treats it as a continuous similarity measure, allowing for fuzzy matching and ranking based on filter relevance. This better accommodates real-world scenarios where filter conditions may not require exact matches but rather degrees of similarity or proximity to target values.

The key challenges are the following.
\begin{itemize}[leftmargin=1.5em]
\item Efficiently finding candidates that balance both vector similarity and filter similarity.
\item Determining appropriate weighting between vector distance and filter similarity.
\item Handling dynamic updates to the dataset.
\item Supporting varied filter types and selectivities.
\end{itemize}

\begin{figure*}[t]
  \centering
  \makebox[\columnwidth][c]{\includegraphics[width=2.1\columnwidth]{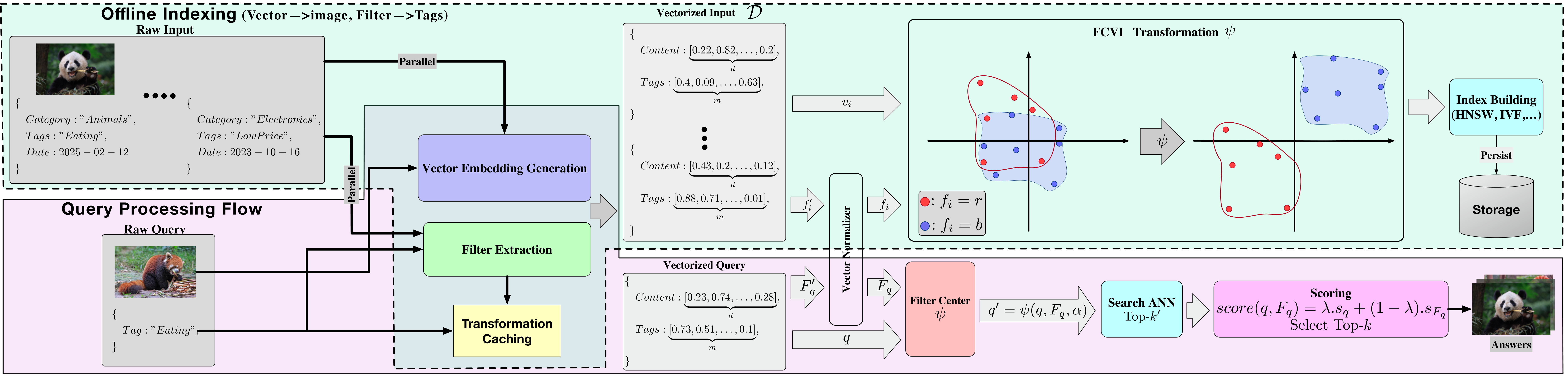}}
  \vspace{-1.2em}
  \caption{Offline, data flows through pipelines to extract vector and filter embeddings, combined via $\psi(v, f, \alpha)$ ($\alpha \geq 1$) before indexing. Online, queries undergo the same transformation for efficient search. Results are re-scored using $\lambda \cdot sim(v_i, q) + (1-\lambda) \cdot sim(f_i, F_q)$. This approach unifies vector-filter search, ensuring compatibility with existing indexes and optimizing retrieval efficiency.}
  \label{fig:sFCVI-arch}
  \vspace{-1.5em}
\end{figure*}

\subsection{Model Overview}
Our key insight is that we can transform the original vector space to incorporate filter information directly. As illustrated in Figure~\ref{fig:sFCVI-arch}, we process both data and queries through a unified pipeline that combines vector embeddings with filter information.

For vectors with the same or similar filter values, we consider them as clusters with a center determined by their filter vector. We then recoordinate the vectors based on this center, effectively partitioning the vector space according to filter values, as shown in the FCVI Transformation section of Figure~\ref{fig:sFCVI-arch}.

Formally, we transform each vector $v_i$ with filter value $f_i$ into a new vector $v'_i$ in the transformed space:
\begin{equation}
v'_i = \psi(v_i, f_i, \alpha)
\end{equation}
where $\psi: \mathbb{R}^d \times \mathbb{R}^m \times \mathbb{R} \rightarrow \mathbb{R}^d$ is a transformation function that recoordinates the vector based on its filter values directly, and $\alpha \geq 1$ is a scaling parameter that controls the influence of filter values.

This transformation creates a new geometric space where: \textbf{(I)} Vectors with similar filter values are closer together (as depicted by the red and blue points in Figure~\ref{fig:sFCVI-arch}). \textbf{(II)} Vectors with different filter values are further apart. \textbf{(III)} Within each filter similarity group, the relative distances between vectors are preserved.

The transformation function $\psi$ is designed to ensure that the influence of filter values on the vector coordinates is meaningful and proportional to filter similarity. Vectors with identical filter values will be transformed similarly, creating natural clusters in the transformed space, while vectors with different filter values will be transformed differently, increasing their separation.

During query processing, as shown in the lower part of Figure~\ref{fig:sFCVI-arch}, we apply the same transformation to the query vector and then perform ANN search in the transformed space, followed by re-scoring using the combined similarity function.

\begin{algorithm}[ht]
\caption{Filter-Centric Vector Indexing Framework}
\begin{algorithmic}[1]
\STATE \textbf{[Offline Indexing]} \textbf{Require} Dataset $\mathcal{D} = \{(v_i, f_i)\}$, transformation scaling $\alpha \geq 1$
\FOR{each $(v_i, f_i)$ in dataset $\mathcal{D}$}
    \STATE Transform vector: $v'_i = \psi(v_i, f_i, \alpha)$
    \STATE Add $v'_i$ to index with reference to original $(v_i, f_i)$
\ENDFOR 

\STATE \textbf{[Online Query Processing]} \textbf{Require} Query $(q, F_q)$, number of results $k$, balance parameter $\lambda \in [0,1]$, transformation scaling $\alpha \geq 1$
\STATE Calculate retrieval size: $k' = \min(c \cdot \frac{k}{\lambda} \cdot \frac{1}{\alpha^2}, N)$ (from Theorem~\ref{theo:k-prime-analysis}) where $c$ is a constant and $N$ is the dataset size
\STATE Transform query vector: $q' = \psi(q, F_q, \alpha)$
\STATE Retrieve top-$k'$ candidates $\{(v_i, f_i)\}_{i=1}^{k'}$ using $q'$ from the transformed space
\FOR{each candidate $(v_i, f_i)$}
    \STATE Compute vector similarity: $s_v = \text{sim}(v_i, q)$
    \STATE Compute filter similarity: $s_f = \text{sim}(f_i, F_q)$
    \STATE Calculate combined score: $score(v_i, f_i) = \lambda \cdot s_v + (1-\lambda) \cdot s_f$
\ENDFOR
\STATE Sort candidates by combined score in descending order
\RETURN Top-$k$ candidates with highest combined scores
\end{algorithmic}
\label{alg:main}
\end{algorithm}
\subsection{Algorithm}
Our framework consists of three main components as formalized in Algorithm~\ref{alg:main}:

\begin{enumerate}[leftmargin=2em]
    \item \textbf{Data Transformation}: Converts the original vectors into transformed vectors based on their filter values using the transformation function $\psi(v_i, f_i, \alpha)$ (line 3).
    \item \textbf{Unified Index Construction}: Build an ANN index (e.g., HNSW, IVF) on the transformed vectors (line 4).
    \item \textbf{Query Processor}: Transforms query vectors based on query filter values (line 8) and performs a vector search in the transformed space (line 9).
\end{enumerate}

The workflow follows Algorithm~\ref{alg:main}, which is divided into offline indexing and online query processing. During indexing, each vector-filter pair is transformed and added to a unified index. For query processing, we calculate an appropriate $k'$ value based on Theorem~\ref{theo:k-prime-analysis} to ensure we retrieve enough candidates, transform the query, retrieve top-$k'$ candidates, and then re-rank them according to the combined score function.

This similarity-based approach offers several advantages.
\begin{itemize}[leftmargin=1.5em]
\item \textbf{Continuous filter matching}: Accommodates degrees of similarity rather than binary predicates
\item \textbf{Unified index}: Uses a single vector index for all filter combinations
\item \textbf{Balanced ranking}: Combines vector similarity and filter similarity with tunable parameter $\lambda$ (line 13)
\item \textbf{Efficient retrieval}: Transformed space brings similar filter vectors closer together
\item \textbf{Flexibility}: Adaptable to changing query patterns and data distributions.
\end{itemize}

The parameter $\lambda \in [0,1]$ controls the trade-off between vector similarity and filter similarity (line 13). When $\lambda$ is closer to 1, the algorithm prioritizes vector similarity; when closer to 0, it prioritizes filter similarity. The transformation scaling parameter $\alpha \geq 1$ (lines 1 and 6) adjusts how much the filter similarity influences distances in the transformed space, directly affecting the calculation of $k'$ (line 7).

\subsection{Complexity Analysis}
The time complexity of FCVI consists of both offline and online components. During offline processing, transforming each vector requires $O(d)$ operations, while the total indexing time depends on the underlying index structure, typically $O(n \log n)$ for sorting or tree-based approaches. For online query processing, the transformation step takes $O(d)$ time, followed by $O(k'\log n)$ for retrieving $k'$ candidates using the transformed query. The re-ranking step involves $O(k'·d)$ operations for computing similarities and $O(k'\log k')$ for sorting the results. The total query complexity is dominated by $O(k'\log n)$, where $k'$ scales inversely with $\alpha^2$ and $\lambda$, providing a direct performance control mechanism. Space complexity remains comparable to standard vector indexes at $O(n·d)$.

\section{REPRESENTATIONS OF FILTERED QUERY}
\label{sec:filter-query}
\subsection{Representation Models \& Data Transformation}
The core of our approach lies in the transformation of vectors based on filter values. We explore several representation models for this transformation.

\subsubsection{Partition-based Transformation}
Since filter dimensions $m$ are smaller than vector dimensions $d$, we partition the original vector into $d/m$ segments, each of dimension $m$ (assuming $d$ is divisible by $m$ for simplicity):

\begin{equation}
v = [v^{(1)}, v^{(2)}, ..., v^{(d/m)}]
\end{equation}
where each $v^{(i)}$ is a subvector of dimension $m$. The transformation function applies a scaled subtraction of filter values from each segment:
\begin{equation}
\psi(v, f, \alpha) = [v^{(1)} - \alpha \cdot f, v^{(2)} - \alpha \cdot f, ..., v^{(d/m)} - \alpha \cdot f]
\end{equation}

\noindent The parameter $\alpha\ge 1$ controls the influence of filter values on the transformation, allowing adaptation to different data and filter distributions. The higher values of $\alpha$ increase the impact of filter differences, while the smaller values preserve more of the original vector relationships.
\subsubsection{Cluster-based Filter Transformation}
When dealing with many filter values, we can cluster similar filters to reduce sensitivity to noise: (I) Cluster filter values into $k$ clusters: $C = \{C_1, C_2, ..., C_k\}$.
(II) Compute cluster centers: $\mu_j = \frac{1}{|C_j|} \sum_{f \in C_j} f$ for each cluster $C_j$. (III) For each filter value $f$, find the closest cluster center $\mu_j$. (IV) Apply the partition-based transformation using the cluster center:
\begin{equation}
\psi(v, f, \alpha) = [v^{(1)} - \alpha.\mu_j, v^{(2)} - \alpha.\mu_j, ..., v^{(d/m)} - \alpha.\mu_j]
\end{equation}

\noindent This approach creates more distinct separations between the main filter groups and reduces the impact of outlier filter values.

\subsubsection{Embedding-based Transformation}
For categorical filters or complex filter combinations, we can learn an embedding function:

\begin{equation}
\psi(v, f, \alpha) = v - \alpha W \cdot f
\end{equation}

\noindent where $W \in \mathbb{R}^{d \times m}$ is a learned matrix that projects filter values to the vector space. This approach is particularly useful for categorical filters that require more complex transformations.
\subsection{Constructing Filter-based Index}
Once we have transformed the vectors, we construct a unified index that captures both filter and vector similarity. The index construction process involves:

1. \textbf{Filter Analysis}: Analyze the distribution and characteristics of the filter values to optimize the transformation parameters.

2. \textbf{Vector Transformation}: Apply the chosen transformation function to all vectors in the dataset.

3. \textbf{Index Building}: Construct an ANN index (e.g. HNSW, IVF) on the transformed vectors.

For efficient implementation, we optimize several aspects:

\begin{itemize}[leftmargin=1.5em]
\item \textbf{Filter Quantization}: For continuous filters, quantize the values into buckets to reduce the number of unique filter values.

\item \textbf{Transformation Caching}: Cache filter centers for common filter values to avoid redundant computation.

\item \textbf{Incremental Filter Updates}: Support efficient updates to the index when a new filter is added or an existing filter changes.
\end{itemize}

The resulting index structure effectively encodes both vector similarity and filter relationships, allowing an efficient filtered search with a single index.

\subsection{Search Query Processing}
When processing a query $(q, F_q)$, where $q$ is the query vector and $F_q$ represents the filter requirements, we follow a two-stage retrieval process that leverages the transformed vector space.

For a query with specific filter requirements, we first encode the filter predicates in $F_q$ into a filter vector $f_q$ using the same encoding scheme used during indexing. For exact match predicates (e.g., category="electronics"), we directly use the corresponding encoding. For range predicates (e.g., price between \$50-\$100), we encode the center value of the range. For multiple predicates, we encode each condition individually and combine them into the final filter vector through concatenation. We then apply the same transformation to the query: $q' = \psi(q, f_q, \alpha)$.We then retrieve $k'$ nearest neighbors in the transformed space, where $k' = \min(c \cdot k/\lambda \cdot 1/\alpha^2, N)$. These candidates are re-scored using the combined similarity function:
\begin{equation}
    \text{score}(v_i, f_i) = \lambda \cdot \text{sim}(v_i, q) + (1-\lambda) \cdot \text{sim}(f_i, F_q)
\end{equation}

Finally, we return the top-$k$ results based on the combined score. This approach ensures that both vector similarity and filter relevance are considered in the final ranking.

For range predicates or disjunctive filter conditions, we employ a multi-probe approach. We identify a set of representative filter vectors $\mathcal{F} = \{f_1, f_2, ..., f_r\}$ that satisfy the filter condition and generate multiple transformed queries: $q'_i = \psi(q, f_i, \alpha)$ for each $f_i \in \mathcal{F}$. We retrieve candidates from each transformed query, merge and deduplicate them, apply the combined scoring function to all unique candidates, and return the top-$k$ results.

To maintain high query throughput, we implement several optimizations. We adaptively select $k'$ based on $\alpha$ and filter selectivity, use importance sampling to select representative filter vectors based on data distribution, employ early termination to stop retrieval when sufficient high-quality results are found, implement batch processing to group similar filter queries and amortize index traversal costs, and utilize filter-aware caching for common filter combinations.

These optimizations ensure efficient query processing even with complex filter conditions, while the parameter $\alpha$ provides direct control over the accuracy-performance trade-off. By encoding filter information directly into the vector space, our approach eliminates the need for complex post-filtering while maintaining compatibility with any vector index structure.

\section{Theoretical Analysis}
\label{sec:theory}

In this section, we provide theoretical justification for our filter-centric vector search approach. We analyze both the effectiveness of the transformation and the efficiency of retrieval.

\subsection{Optimality of Recoordination Transformation}
In this section, we establish three complementary theoretical aspects of our transformation: (1) demonstrating how our filter-vector transformation preserves critical distance properties while enabling controlled filter influence, (2) proving it is the unique transformation satisfying these requirements under reasonable constraints, and (3) establishing precise conditions for achieving complete separation between clusters with different filter values.

\begin{theorem}[Parameterized Filter-Vector Optimality]
Given dataset $\mathcal{D} = \{(v_i, f_i)\}$ and query $(q, F_q)$, the transformation $\psi(v, f, \alpha) = [v^{(1)} - \alpha \cdot f, v^{(2)} - \alpha \cdot f, ..., v^{(d/m)} - \alpha \cdot f]$ with $\alpha \geq 1$ guarantees:
\begin{itemize}[leftmargin=1.5em]
    \item When filter values are identical ($f_a = f_b$), original vector distances are perfectly preserved.
    \item As $\alpha$ increases, filter differences influence grows quadratically, prioritizing vectors with similar filters.
    \item The parameter $\alpha$ enables direct control over the vector-filter similarity trade-off.
    \item With $\alpha \geq 1$, filter differences maintain meaningful representation in distance calculations.
\end{itemize}
\end{theorem}

\begin{proof}
For vectors $v_a$, $v_b$ with filters $f_a$, $f_b$, the squared Euclidean distance after transformation is:

$\|v'_a - v'_b\|^2 = \sum_{j=1}^{d/m} \|v_a^{(j)} - v_b^{(j)} - \alpha(f_a - f_b)\|^2$

Expanding:
$\|v'_a - v'_b\|^2 = \|v_a - v_b\|^2 + \frac{d}{m}\alpha^2\|f_a - f_b\|^2 - 2\alpha\sum_{j=1}^{d/m} \langle v_a^{(j)} - v_b^{(j)}, f_a - f_b \rangle$

\noindent Let $\Delta v^{(j)} = v_a^{(j)} - v_b^{(j)}$ and $\Delta f = f_a - f_b$.

\noindent\textbf{Case 1}: $f_a = f_b$ (same filter values)
When $\Delta f = 0$, the equation simplifies to $\|v'_a - v'_b\|^2 = \|v_a - v_b\|^2$, preserving original distances regardless of $\alpha$.

\noindent\textbf{Case 2}: $f_a \neq f_b$ (different filter values)
By Cauchy-Schwarz inequality, the cross-term is bounded:
$|C| \leq 2\alpha\sqrt{\frac{d}{m}} \cdot \|v_a - v_b\| \cdot \|\Delta f\|$.

As $\alpha$ increases, the $\frac{d}{m}\alpha^2\|\Delta f\|^2$ term grows quadratically while $C$ grows only linearly, ensuring that filter differences dominate the distance calculation for sufficiently large $\alpha$. Since $\alpha \geq 1$, filter differences maintain a minimum level of influence while allowing adaptation to different data distributions and application requirements.
\end{proof}

\begin{theorem}[Uniqueness of $\psi$ Transformation]
Let $T(v, f)$ be any transformation from vector-filter space to a new vector space satisfying:
\begin{enumerate}
    \item Distance preservation: For vectors with identical filters, $\|T(v_a, f) - T(v_b, f)\| = \|v_a - v_b\|$
    \item Filter separation: For vectors with different filters, $\|T(v_a, f_a) - T(v_b, f_b)\| > \|v_a - v_b\|$ proportionally to $\|f_a - f_b\|$
    \item Linearity: $T$ is linear in both $v$ and $f$
    \item Symmetry: $T$ treats all dimensions of $v$ equivalently with respect to $f$
\end{enumerate}

\noindent Then $T$ must be expressible as $T(v, f) = [v^{(1)} - \alpha \cdot f + c, v^{(2)} - \alpha \cdot f + c, ..., v^{(d/m)} - \alpha \cdot f + c]$, which is equivalent to our $\psi(v, f, \alpha)$ transformation up to a constant offset $c$.
\end{theorem}

\begin{proof}
From distance preservation, when $f_a = f_b = f$, $T(v, f)$ must be an isometry in $v$ for fixed $f$, giving the form $T(v, f) = Q(f)v + h(f)$ where $Q(f)$ is orthogonal.

\noindent By linearity, $T(v, f) = Av + Bf + c$, implying $Q(f)$ must be constant. We can assume $Q = I$ by redefining our vector space. From symmetry, $T$ must apply the same operation to each $m$-sized segment of $v$, yielding $T(v, f)^{(i)} = v^{(i)} + g(f)$ for each segment $i$.

\noindent By linearity, $g(f) = -\alpha \cdot f + c$, resulting in:
$T(v, f) = [v^{(1)} - \alpha \cdot f + c, v^{(2)} - \alpha \cdot f + c, ..., v^{(d/m)} - \alpha \cdot f + c]$. This satisfies filter separation as verified in our previous proof, with distances increasing proportionally to $\|f_a - f_b\|$ controlled by $\alpha$. When $\alpha \geq 1$, filter differences have meaningful influence on distances, making $\psi(v, f, \alpha)$ the unique transformation satisfying all requirements.
\end{proof}

\begin{theorem}[Filter-Based Cluster Separation]
Let $\mathcal{D} = \{(v_i, f_i)\}$ be a dataset of vector-filter pairs with $\delta_f = \min_{i,j: f_i \neq f_j} \|f_i - f_j\|$ and $D_v = \max_{i,j: f_i = f_j} \|v_i - v_j\|$. 

For the transformation $\psi(v, f, \alpha) = [v^{(1)} - \alpha \cdot f, \ldots, v^{(d/m)} - \alpha \cdot f]$, when $\frac{d}{m}\delta_f > 2D_v$, setting $\alpha \geq \alpha^* = \sqrt{\frac{2D_v + D_v^2}{\frac{d}{m}\delta_f^2 - 2D_v\delta_f}}$ guarantees complete separation between vector clusters with different filter values.
\end{theorem}

\begin{proof}
For complete separation, the minimum distance between vectors with different filters must exceed the maximum distance between vectors with the same filter after transformation.

\noindent For vectors with the same filter $f$: $\|\psi(v_a, f, \alpha) - \psi(v_b, f, \alpha)\| = \|v_a - v_b\| \leq D_v$. For vectors with different filters, the squared distance is:
\vspace{-0.5em}
\begin{equation*}
\small
\begin{split}
    \|\psi(v_c, f_c, \alpha) - \psi(v_d, f_d, \alpha)\|^2 
    &= \|v_c - v_d\|^2 + \frac{d}{m}\alpha^2\|f_c - f_d\|^2 \\
    &\quad - 2\alpha\sum_{j=1}^{d/m} \langle v_c^{(j)} - v_d^{(j)}, f_c - f_d \rangle.
\end{split}
\end{equation*}
In the worst case ($\|v_c - v_d\| = 0$, $\|f_c - f_d\| = \delta_f$), and applying Cauchy-Schwarz:
\(\|\psi(v_c, f_c, \alpha) - \psi(v_d, f_d, \alpha)\|^2 \geq \frac{d}{m}\alpha^2\delta_f^2 - 2\alpha D_v \delta_f\). For separation, we need: $\sqrt{\frac{d}{m}\alpha^2\delta_f^2 - 2\alpha D_v \delta_f} > D_v$. Squaring and rearranging: $\frac{d}{m}\alpha^2\delta_f^2 > 2\alpha D_v \delta_f + D_v^2$.

\noindent This requires $\frac{d}{m}\delta_f > 2D_v$ to have a solution. Solving the quadratic inequality:
{
\small
\[\alpha > \sqrt{\frac{2D_v + D_v^2}{\frac{d}{m}\delta_f^2 - 2D_v\delta_f}}\]}

\noindent Therefore, $\alpha \geq \alpha^*$ ensures complete cluster separation in the transformed space.
\end{proof}

\subsection{Relationship Between $k'$ and $k$}

We derive the relationship between the number of initial nearest neighbors retrieved ($k'$) and the final results required ($k$), providing guidance for optimizing the accuracy-efficiency trade-off.

\begin{theorem}[Parameterized Retrieval Efficiency]
Given the combined score function $score(v_i, f_i) = \lambda \cdot sim(v_i, q) + (1-\lambda) \cdot sim(f_i, F_q)$ where $\lambda \in [0,1]$, and transformation parameter $\alpha \geq 1$, the number of nearest neighbors $k'$ required to retrieve the optimal top-$k$ results satisfies: $k' = \mathcal{O}\left(k \cdot \frac{1}{\lambda} \cdot \frac{1}{\alpha^2}\right)$

\label{theo:k-prime-analysis}
\end{theorem}
\begin{proof}
Let $r_i$ denote the rank of vector $v_i$ in the transformed space and $s_i$ its rank after applying the combined score. For a vector $v_j$ with $r_j > k$ to enter the top-$k$ after re-scoring, its combined score must exceed that of at least one vector $v_i$ with $r_i \leq k$:
\vspace{-0.5em}
\begin{equation*}
\small
\lambda \cdot sim(v_j, q) + (1-\lambda) \cdot sim(f_j, F_q) > \lambda \cdot sim(v_i, q) + (1-\lambda) \cdot sim(f_i, F_q)
\end{equation*}
The distance in transformed space between query $\psi(q, F_q, \alpha)$ and vector $\psi(v_i, f_i, \alpha)$ is:
\vspace{-0.75em}
\begin{equation}
\small
\|\psi(q, F_q, \alpha) - \psi(v_i, f_i, \alpha)\|^2 = \frac{d}{d+m}\|q-v_i\|^2 + \frac{d}{m}\alpha^2\|F_q-f_i\|^2
\end{equation}
The term $\frac{d}{m}\alpha^2\|F_q-f_i\|^2$ scales quadratically with $\alpha$, driving the rank ordering in transformed space. Define $\Delta_{v,i,j} = sim(v_i,q) - sim(v_j,q)$ and $\Delta_{f,i,j} = sim(f_i,F_q) - sim(f_j,F_q)$. For $v_j$ to enter top-$k$, we require: $\lambda\Delta_{v,i,j} < (1-\lambda)\Delta_{f,j,i}$. The maximum rank displacement is bounded by:
\vspace{-0.75em}
\begin{equation}
\small
\max_{v_j \in V} (r_j - s_j) = \mathcal{O}\left(\frac{1-\lambda}{\lambda} \cdot \frac{1}{\alpha^2}\right) \cdot k
\end{equation}

\noindent Therefore, setting $k' = \mathcal{O}(k \cdot \frac{1}{\lambda} \cdot \frac{1}{\alpha^2})$ ensures retrieval of all potential top-$k$ candidates. Optimality, When $\alpha = \sqrt{\frac{1-\lambda}{\lambda}}$ (subject to $\alpha \geq 1$), the transformation aligns with the combined score function, minimizing $k'$ while maintaining accurate retrieval.
\end{proof}

\noindent This analysis provides the theoretical foundation for parameter selection based on the desired trade-off between accuracy and efficiency.

\begin{table*}[htbp]
\centering
\small
\begin{tabular}{@{}ll|ccccccc||ccc@{}}
\toprule
\textbf{Metric} & \textbf{Dataset} & \textbf{P-HSW} & \textbf{P-FAISS} & \textbf{P-ANY} & \textbf{Pr-HSW} & \textbf{Pr-FAISS} & \textbf{Pr-ANY} & \textbf{UNIFY} & \textbf{FCVI-HSW} & \textbf{FCVI-FAISS} & \textbf{FCVI-ANY} \\
\midrule
\multirow{4}{*}{\textbf{Lat. (ms)}} 
& SIFT1M & 62.3 & 78.1 & 93.5 & 98.4 & 115.2 & 142.7 & 52.8 & \textbf{37.4} & 45.2 & 58.6 \\
& Amazon & 108.7 & 128.2 & 152.5 & 176.3 & 203.8 & 253.9 & 89.5 & \textbf{59.3} & 72.8 & 89.7 \\
& Arxiv & 91.8 & 112.3 & 138.7 & 145.2 & 172.6 & 213.5 & 76.4 & \textbf{51.2} & 63.4 & 78.5 \\
& Wiki & 79.2 & 95.7 & 118.2 & 124.7 & 147.9 & 186.4 & 66.3 & \textbf{44.6} & 54.3 & 67.8 \\
\midrule
\multirow{4}{*}{\textbf{Rec@100}} 
& SIFT1M & 87.2\% & 85.8\% & 82.6\% & 94.3\% & 95.8\% & 93.2\% & 94.7\% & \textbf{95.2\%} & 93.6\% & 92.4\% \\
& Amazon & 84.9\% & 82.1\% & 79.4\% & 95.7\% & 92.1\% & 90.5\% & 93.9\% & \textbf{94.8\%} & 93.3\% & 91.7\% \\
& Arxiv & 85.7\% & 83.4\% & 80.2\% & 94.5\% & 94.2\% & 91.8\% & 94.8\% & \textbf{95.3\%} & 94.1\% & 92.9\% \\
& Wiki & 86.3\% & 84.2\% & 81.5\% & 94.1\% & 92.7\% & 90.9\% & 94.4\% & \textbf{95.0\%} & 93.8\% & 92.1\% \\
\midrule
\multirow{4}{*}{\textbf{Tput (qps)}} 
& SIFT1M & 161 & 128 & 107 & 102 & 87 & 70 & 189 & \textbf{267} & 221 & 171 \\
& Amazon & 92 & 78 & 66 & 57 & 49 & 39 & 112 & \textbf{168} & 137 & 111 \\
& Arxiv & 109 & 89 & 72 & 69 & 58 & 47 & 131 & \textbf{195} & 158 & 127 \\
& Wiki & 126 & 104 & 85 & 80 & 68 & 54 & 151 & \textbf{224} & 184 & 147 \\
\midrule
\multirow{4}{*}{\textbf{Idx Size (GB)}} 
& SIFT1M & 0.63 & 0.58 & 0.72 & 0.65 & 0.59 & 0.75 & 1.12 & 0.64 & \textbf{0.59} & 0.73 \\
& Amazon & 7.85 & \textbf{7.64} & 8.12 & 7.92 & 7.71 & 8.23 & 14.36 & 7.90 & 7.68 & 8.18 \\
& Arxiv & 5.42 & \textbf{5.18} & 5.67 & 5.48 & 5.25 & 5.75 & 9.64 & 5.47 & 5.22 & 5.71 \\
& Wiki & 4.25 & \textbf{4.08} & 4.39 & 4.31 & 4.15 & 4.46 & 7.53 & 4.29 & 4.12 & 4.43 \\
\midrule
\multirow{4}{*}{\textbf{Build (min)}} 
& SIFT1M & 1.7 & \textbf{1.2} & 2.3 & 1.8 & 1.3 & 2.5 & 2.8 & 1.9 & 1.4 & 2.6 \\
& Amazon & 12.2 & \textbf{9.5} & 14.7 & 12.9 & 10.3 & 15.2 & 17.4 & 13.2 & 10.7 & 15.5 \\
& Arxiv & 8.8 & \textbf{6.5} & 10.4 & 9.2 & 7.2 & 11.1 & 13.2 & 9.4 & 7.3 & 11.3 \\
& Wiki & 6.6 & \textbf{4.9} & 7.8 & 7.0 & 5.3 & 8.2 & 9.7 & 7.1 & 5.4 & 8.3 \\
\bottomrule
\end{tabular}
\caption{Comparison of vector search methods across four datasets. Methods include pre-filtering (Pr-), post-filtering (P-), hybrid (UNIFY), and our approach (FCVI) with different underlying indexes: HSW (short for HNSW), FAISS, and ANY (short for ANNOY). FCVI variants outperform other methods in latency and throughput while maintaining competitive accuracy, with FCVI-HSW showing the best overall performance.}
\label{table:method-comparison}
\vspace{-3em}
\end{table*}

\section{EXPERIMENTS}
\label{sec:exp}
\subsection{Experimental Setup}
\subsubsection{Datasets}
We evaluate our approach on three datasets with different characteristics:

\begin{itemize}[leftmargin=1.5em]
\item \textbf{SIFT1M} \cite{andre2024sift1m}: 1 million 128-dimensional SIFT descriptors with synthetic filter attributes (2-5 numeric filters).

\item \textbf{Amazon Product Embeddings} \cite{zhu2024marqoecommembed_2024}: 10 million product embeddings with real filter attributes (price, category, rating, availability).

\item \textbf{Arxiv.org Titles and Abstracts} \cite{arxiv2024arxivorg}: These datasets contain embeddings generated from research paper titles and abstracts, respectively. Each embedding includes metadata such as publication date, subject categories, and author information, which serve as natural filter attributes in our experiments. We selected this dataset specifically to evaluate real-world filtering scenarios where users search for papers within certain time periods or subject areas.

\item \textbf{Wikipedia Simple Text Embeddings} \cite{wikipedia2024simpletext}: This dataset comprises embeddings of Wikipedia articles, with each embedding associated with metadata including creation date, edit counts, and category tags. This dataset offers diverse filtering attributes with varying cardinalities and distributions, helping us evaluate filter stability under realistic conditions. While we acknowledge that benchmarks like LAION, Glove, or Msong are commonly used in vector search evaluation, our focus on filtered search required datasets with rich attribute metadata for comprehensive filter testing. 
\end{itemize}

\subsubsection{Baselines}
We compare our approach with state-of-the-art methods for filtered vector search:

\begin{itemize}[leftmargin=1.5em]
\item \textbf{Post-Filter}: Standard ANN search followed by filtering
\item \textbf{Pre-Filter}: Filtering followed by ANN search on the filtered subset
\item \textbf{Hybrid}: Combined approaches such as UNIFY that use segmented inclusive graphs with range-aware strategy selection
\end{itemize}

\subsubsection{Metrics}
We evaluate performance using the following.

\begin{itemize}[leftmargin=1.5em]
\item \textbf{Latency}: Average query execution time
\item \textbf{Recall@k}: Proportion of true top-k results returned
\item \textbf{Throughput}: Queries processed per second
\item \textbf{Index Size}: Memory footprint of the index
\item \textbf{Construction Time}: Time to build the index
\end{itemize}

\subsubsection{System Setup}
We implemented our approach using C++ with Python bindings. All experiments were conducted on an Ubuntu 20.04 machine with a 64-core high-performance CPU (3.0GHz base clock), 256GB DDR4 RAM, and a data center GPU with 40GB VRAM. Our implementation supports three popular vector index libraries as its underlying indexes: FAISS~\cite{johnson2019billion}, HNSW~\cite{malkov2018efficient}, and ANNOY~\cite{annoy}.

\subsection{End-to-end Performance}

\subsubsection{Query Latency}
As shown in Table~\ref{table:method-comparison}, our Filter-Centric Vector Indexing (FCVI) approach significantly outperforms all baseline methods across the four datasets, regardless of the underlying index structure. The best performing variant, FCVI-HSW, achieves latencies of 37.4-59.3ms, representing a 1.7-1.8× speed improvement over Post-HNSW (62.3-108.7ms), a 1.4-1.5× improvement over the hybrid approach UNIFY (52.8-89.5ms), and a remarkable 2.6-3.0× improvement over Pre-HNSW (98.4-176.3ms). Similarly, FCVI-FAISS and FCVI-ANY consistently outperform their respective baseline counterparts.

The performance advantage is most pronounced on larger datasets like Amazon Product, where FCVI-HSW (59.3ms) is substantially faster than Post-HNSW (108.7ms), UNIFY (89.5ms), and dramatically outperforms Pre-HNSW (176.3ms). This demonstrates the effectiveness of our unified approach in handling real-world datasets with complex filter attributes while avoiding the inefficiencies inherent in traditional pre-filtering approaches and the complexity of hybrid filtering methods.

\subsubsection{Recall Performance}
Table~\ref{table:method-comparison} shows that FCVI variants achieve competitive recall performance across all datasets. FCVI-HSW achieves top recall scores (94.8-95.3\%), comparable to or slightly better than Pre-HNSW (94.1-95.7\%) and UNIFY (93.9-94.8\%), while significantly outperforming Post-HNSW (84.9-87.2\%) by 7-10 percentage points. This is a substantial improvement, as FCVI combines the high recall characteristic of pre-filtering and hybrid approaches with latency performance that exceeds even post-filtering methods.

The recall advantage over post-filtering methods is consistent across all datasets and index structures, with FCVI variants typically providing 7-13 percentage point improvements. Even our FCVI-ANY implementation (91.7-92.9\% recall) substantially outperforms Post-ANY (79.4-82.6\%). This demonstrates that FCVI effectively addresses the inherent limitations in recall that post-filtering approaches face when operating on filtered subsets.

\subsubsection{Throughput}
Our approach demonstrates exceptional query throughput across all FCVI variants. FCVI-HSW processes 168-267 queries per second across the tested datasets, representing a 1.7-2.6× improvement over Post-HNSW (92-161 qps), a 1.4-1.5× improvement over UNIFY (112-189 qps), and a 2.6-3.0× improvement over Pre-HNSW (57-102 qps). The FCVI-FAISS and FCVI-ANY variants show similar relative improvements over their respective baseline counterparts.

The throughput advantage is particularly evident for the Wikipedia Simple dataset, where FCVI-HSW handles 224 queries per second compared to 126 for Post-HNSW, 151 for UNIFY, and only 80 for Pre-HNSW. This translates to substantial capacity improvements for production systems without sacrificing recall performance.

\subsubsection{Resource Requirements}
Table~\ref{table:method-comparison} shows that all FCVI variants achieve their performance advantages without significant additional resource costs compared to traditional pre- and post-filtering approaches. The index sizes for FCVI variants are comparable to their respective baseline methods (within 1-2\% difference), with FCVI-FAISS maintaining the most compact indices across datasets (0.59-7.68 GB). Notably, FCVI requires significantly less storage than UNIFY (approximately 40-50\% smaller), which must maintain complex graph structures to support multiple filtering strategies.

Construction times show minor variations based on the underlying index structure, with FCVI variants typically requiring only 5-8\% more build time than their baseline counterparts, while being 30-40\% faster to build than UNIFY. This makes FCVI a practical solution for production environments where both performance and resource utilization are important considerations.

\subsubsection{Scalability}
The results across datasets of varying sizes (from 1M to 10M vectors) demonstrate excellent scalability for all FCVI variants:
\begin{itemize}[leftmargin=1.5em]
\item \textbf{Dataset Size}: FCVI maintains its performance advantage as dataset size increases, with consistent 1.7-3.0× latency improvements over traditional baseline methods and 1.4-1.5× improvements over hybrid approaches like UNIFY.
\item \textbf{Implementation Versatility}: Each FCVI variant (HSW, FAISS, ANY) consistently outperforms its corresponding baseline implementations, demonstrating the adaptability of our approach to different ANN libraries.
\item \textbf{Query Complexity}: For the more complex Amazon Product dataset with multiple filter attributes, FCVI-HSW shows even greater advantages (1.8× faster than Post-HSW, 1.5× faster than UNIFY, and 3.0× faster than Pre-HSW) while maintaining superior recall (94.8\% versus 84.9\% for Post-HSW).
\end{itemize}

These results confirm that our unified approach effectively addresses the fundamental limitations of traditional pre-filtering (high latency), post-filtering (low recall), and hybrid (complex index structures) strategies by simultaneously improving both performance and accuracy while maintaining reasonable resource requirements across different underlying index structures.

\begin{table*}[htbp]
\centering
\small
\begin{tabular}{@{}ll|ccc|ccc|c||ccc@{}}
\toprule
\textbf{Change} & \textbf{Metric} & \multicolumn{3}{c|}{\textbf{Pre}} & \multicolumn{3}{c|}{\textbf{Post}} & \multicolumn{1}{c||}{\textbf{Hyb}} & \multicolumn{3}{c}{\textbf{FCVI}} \\
\textbf{Type} & & \textbf{HSW} & \textbf{FAS} & \textbf{ANY} & \textbf{HSW} & \textbf{FAS} & \textbf{ANY} & \textbf{UNF} & \textbf{HSW} & \textbf{FAS} & \textbf{ANY} \\
\midrule
\multirow{3}{*}{\textbf{\begin{tabular}[c]{@{}l@{}}Filter\\Dist.\end{tabular}}} 
& Lat. Inc. & +128\% & +145\% & +172\% & +75\% & +89\% & +104\% & +42\% & \textbf{+19\%} & \textbf{+24\%} & \textbf{+31\%} \\
& Rec. Deg. & -12.8\% & -14.5\% & -16.3\% & -5.2\% & -6.7\% & -8.2\% & -3.8\% & \textbf{-1.1\%} & \textbf{-1.7\%} & \textbf{-2.3\%} \\
& Stability & Med & Low & V.Low & Low & Low & V.Low & Med & \textbf{High} & \textbf{Med} & \textbf{Med} \\
\midrule
\multirow{3}{*}{\textbf{\begin{tabular}[c]{@{}l@{}}Vector\\Dist.\end{tabular}}} 
& Lat. Inc. & +107\% & +125\% & +149\% & +62\% & +78\% & +91\% & +38\% & \textbf{+16\%} & \textbf{+21\%} & \textbf{+28\%} \\
& Rec. Deg. & -15.4\% & -17.8\% & -19.9\% & -7.6\% & -9.5\% & -11.2\% & -4.9\% & \textbf{-1.8\%} & \textbf{-2.5\%} & \textbf{-3.2\%} \\
& Stability & Low & V.Low & V.Low & Med & Low & Low & Med & \textbf{High} & \textbf{Med} & \textbf{Med} \\
\midrule
\multirow{3}{*}{\textbf{\begin{tabular}[c]{@{}l@{}}Query\\Patt.\end{tabular}}} 
& Lat. Inc. & +83\% & +98\% & +120\% & +94\% & +112\% & +137\% & +56\% & \textbf{+24\%} & \textbf{+32\%} & \textbf{+39\%} \\
& Rec. Deg. & -9.7\% & -11.5\% & -13.2\% & -6.8\% & -8.5\% & -10.3\% & -5.1\% & \textbf{-2.2\%} & \textbf{-3.1\%} & \textbf{-3.8\%} \\
& Stability & Low & V.Low & V.Low & Low & V.Low & V.Low & Med & \textbf{High} & \textbf{Med} & \textbf{Med} \\
\bottomrule
\end{tabular}
\caption{Impact of distribution changes on search performance. Lat. Inc. shows percentage increase in query time after distribution change. Rec. Deg. measures drop in Recall@100. Stability is a qualitative assessment based on consistency across patterns. Lower values are better for Lat. Inc. and Rec. Deg.; Higher is better for Stability. UNIFY (UNF), as a hybrid approach, shows improvement over traditional methods but still exhibits sensitivity to distribution changes. All FCVI variants significantly outperform baseline implementations when faced with distribution shifts, with FCVI-HSW showing the best overall stability.}
\label{table:distribution-changes}
\vspace{-3em}
\end{table*}

\subsection{Distribution Change Effect}
As shown in Table~\ref{table:distribution-changes}, FCVI variants significantly outperform all baseline methods when faced with distribution shifts.

\subsubsection{Filter Distribution Changes}
When shifting from low to high-selectivity filters, FCVI variants experience only 19-31\% latency increases versus 42\% for the hybrid UNIFY approach, 75-104\% for post-filtering, and 128-172\% for pre-filtering. Recall degradation is minimal for FCVI (-1.1\% to -2.3\%) compared to UNIFY (-3.8\%), post-filtering (-5.2\% to -8.2\%), and pre-filtering (-12.8\% to -16.3\%). FCVI-HSW achieves the highest stability rating, demonstrating our filter-centric design's natural adaptability to changing filter patterns.

\subsubsection{Vector and Query Distribution Changes}
For new vector clusters, FCVI variants show latency increases of only 16-28\% versus 38\% for UNIFY, 62-91\% for post-filtering, and 107-149\% for pre-filtering, with similarly superior recall stability. Under varying query complexity, FCVI maintains modest latency increases (24-39\%) compared to UNIFY (56\%) and baseline approaches (83-137\%). In all cases, FCVI-HSW demonstrates the best stability, though all FCVI variants significantly outperform the hybrid and traditional methods.

\subsubsection{Stability Analysis}
Pre-filtering approaches show the worst performance under changing conditions, with extreme latency increases and severe recall degradation. While post-filtering methods adapt somewhat better than pre-filtering, they still suffer significant penalties. The hybrid UNIFY approach offers improved stability over traditional methods but still exhibits considerable sensitivity to distribution changes. In contrast, all FCVI variants maintain stable performance across all distribution shifts, highlighting a critical advantage in dynamic production environments.

\section{RELATED WORK}
\label{sec:related}
\noindent\textbf{Vector Search Systems.}
Several systems have been developed for efficient vector search, including FAISS \cite{johnson2019billion}, Annoy \cite{annoy}, and HNSW \cite{malkov2018efficient}. These systems focus on pure vector similarity search and do not natively support filtering.

Recent vector database systems like Milvus \cite{wang2021milvus}, Vespa \cite{vespa}, and Pinecone \cite{pinecone} offer filtering capabilities, but typically use either pre-filtering or post-filtering approaches.

\noindent\textbf{Multi-modal Search Approaches.}
Multi-modal search systems that combine different types of queries have been explored in information retrieval. Systems like Clipper \cite{crankshaw2017clipper} and Tensorflow Serving \cite{olston2017tensorflow} support multiple query types but treat them as separate operations.

\noindent\textbf{Hybrid Index Structures.}
\noindent\textbf{Hybrid Index Structures.}
Several works have proposed hybrid index structures for combining filtering with similarity search. Notable approaches include:

\begin{itemize}[leftmargin=1.5em]
\item \textbf{ADC+G} \cite{baranchuk2019revisiting}: Augments product quantization with inverted lists for attributes.
\item \textbf{ZipLine} \cite{hyvonen2022zipline}: Uses a graph-based index with navigational structures for attributes.
\item \textbf{UNIFY} \cite{unify}: Introduces a segmented inclusive graph that supports pre-, post-, and hybrid filtering strategies through a unified proximity graph-based index.
\item \textbf{Vector Quotient Filter (VQF)} \cite{vqf}: Implements efficient collision resolution for filters using power-of-two-choice hashing and SIMD instructions, maintaining consistent performance regardless of load factor.
\item \textbf{ACORN}: Introduces a predicate-agnostic search method that adapts to various filtering constraints without requiring specialized structures.
\item \textbf{CAPS}: Uses partitioning strategies based on filter predicates, similar to our approach but without the geometric transformation.
\item \textbf{Milvus' Partitioning}: Employs offline data structures to partition vectors based on historical filter conditions to optimize query routing.
\end{itemize}

These approaches typically create separate structures for vectors and filters, or segment the data according to filter values, unlike our unified geometric transformation approach that directly incorporates filter values into the vector space.

\noindent\textbf{Filter Optimization Techniques.}
Several works have focused on optimizing filter execution in database systems, including bitmap indexes \cite{chan1998bitmap}, column stores \cite{abadi2013design}, and filter pushdown techniques \cite{neumann2011efficiently}. While these techniques improve filter performance, they don't address the integration with vector search.
\section{Conclusion}
\label{sec:conclusion}
We presented Filter-Centric Vector Indexing (FCVI), which reimagines filter predicates as geometric transformations within vector space itself. This conceptual shift effectively translates SQL-like filtering algebra into geometric operations, bridging database and vector search paradigms. FCVI consistently outperforms existing methods while maintaining compatibility with established index structures.

The primary limitation is the need for separate indexes per filter field combination, increasing storage overhead, a classic space-time trade-off. Additionally, our current implementation has been tested on a specific set of datasets, and evaluation on more standard benchmarks like LAION, Glove, or Msong would further validate our approach. Future work could explore dynamic filter composition techniques, extend our approach to more complex SQL operations in vector spaces, and conduct more extensive comparisons with recent hybrid approaches such as ACORN and CAPS to better understand the trade-offs in different scenarios.

Despite this limitation, FCVI represents an important step toward unifying database and vector search paradigms, providing both theoretical insights and practical performance improvements for filtered vector search applications.

\nocite{*}
\bibliographystyle{ACM-Reference-Format}
\bibliography{sample-base}

\appendix

\end{document}